\documentclass[psfig,useAMS,usenatbib]{mn2e}
\usepackage{aas_macros}
\usepackage{psfig}
\usepackage{graphicx}
\usepackage{epstopdf}
\usepackage{bm}

\date{Accepted ---. Received ---; in original form ---.}

\begin{document}

\title{Cosmic Shear Systematics: Software-Hardware Balance}
\author[A. Amara, A. R{\'e}fr{\'e}gier, S. Paulin-Henriksson]
{A. Amara$^{1}$, A. R{\'e}fr{\'e}gier$^{2}$, S. Paulin-Henriksson$^{2}$\\
$^{1}$Department of Physics, ETH Zurich, Wolfgang-Pauli-Strasse 16,
  CH-8093 Zurich, Switzerland\\
$^{2}$Service d'Astrophysique, CEA Saclay, 91191 Gif sur Yvette, France.}

\maketitle

\begin{abstract}

Cosmic shear measurements rely on our ability to measure and correct the Point Spread Function (PSF) of the observations. This PSF is measured using stars in the field, which give a noisy measure at random points in the field. Using Wiener filtering, we show how errors in this PSF correction process propagate into shear power spectrum errors. This allows us to test future space-based missions, such as Euclid or JDEM, thereby allowing us to set clear engineering specifications on PSF variability. For ground-based surveys, where the variability of the PSF is dominated by the environment, we briefly discuss how our approach can also be used to study the potential of mitigation techniques such as correlating galaxy shapes in different exposures. To illustrate our approach we show that for a Euclid-like survey to be statistics limited, an initial pre-correction PSF ellipticity power spectrum, with a power-law slope of -3 must have an amplitude at $\ell =1000 $ of less than $2\times 10^{-13}$. This is 1500 times smaller than the typical lensing signal at this scale. We also find that the power spectrum of PSF size ($\delta_{R^2}$) at this scale must be below $2\times 10^{-12}$. \\
Public code available as part of \tt{iCosmo} at \tt{http://www.icosmo.org}

\end{abstract}

\section{Introduction}
High accuracy cosmic shear measurements require a precise measure of galaxy shapes, which become correlated due to gravitational lensing.  To measure this lensing induced correlation, we must first correct for any instrument effects that also cause the observed galaxy shapes to correlate.  In particular, the Point Spread function (PSF) of the instrument needs to be corrected. This PSF includes effects associated with the instrument as well as the observing environment such as the atmosphere.  Given an observing system with a given PSF, correcting a galaxy image is done by measuring the PSF using stars in the neighborhood of the galaxy and then deconvolving this from the galaxy image.  In so doing, the two point correlation function of galaxy ellipticities (which is a combination of true galaxy ellipticity and PSF ellipticity correlation functions) can be reduced to true galaxy ellipticity with some residual error.  In \cite{2008MNRAS.391..228A} we showed that future ambitious surveys will need to control these residuals to a level where their contribution variance over the range of scales being used is $\sigma^2_{sys} < 10^{-7}$, where $\sigma^2_{sys}$ is defined in \cite{2008MNRAS.391..228A}.  This results was also confirmed by  \cite{2008arXiv0812.1966K}.

Achieving this will require both tight controls of the inherent PSF correlations of the instrument as well as an accurate method for PSF correction.  In \cite{2008A&A...484...67P} we focused exclusively on the limits of PSF correction without considering the impact of the initial PSF pattern.  This allowed us to investigate the minimal amount of information that we would need to collect in order to correct the PSF so that the residuals stay below $\sigma^2_{sys} < 10^{-7}$.  Since each star gives a finite amount of information about the PSF due to noise and pixelisation, the PSF must be measured by collecting information from a number of stars.  The length scale that corresponds to this minimal number of stars ( $\theta_{min}$) then becomes a key scale, where correlation information of larger scales is `safe' but ellipticity correlations on scales smaller than this are `not safe' unless extra information is known about the PSF.  In this paper, we extend the earlier work in \cite{2008A&A...484...67P} , using Wiener filtering, to include the requirements that would need to be placed on the PSF correlation function if we wish to use scales smaller than $\theta_{min}$.  In this way the design of future surveys can be divided into two regimes: (i) large scales where the adverse affects of PSF are correct in software using image processing; and (ii) small scales that need to be controlled in hardware by, for instance, designing a space-based instrument with a well behaved PSF on small scales.

This paper is organised as follows: Section 2 sets out the basic statistical properties of a field that is sampled at random points. In section 3, we show the impact of using a Wiener filter to model the PSF which leaves residual errors. In section 4, we show the impact for cosmic shear surveys, and finally we discuss our interpretation and conclusions in section 5.

\section{Random Sampling}
To correct an image for the effects of the PSF we must be able to construct a model for the PSF at the desired point, which is usually the position of a galaxy.  We do this by measuring the PSF, which we assume is a continuos function $f(\bm{\theta})$, at discrete points $\bm{\theta}_i$ where the stars lie in the image. We can construct a representation of the observed field $f_{obs}(\bm{\theta})$:
\begin{equation}
f_{obs}(\bm{\theta}) = f(\bm{\theta})\eta(\bm{\theta}) + \varepsilon(\bm{\theta})\eta(\bm{\theta}),
\label{eq:dis_samp}
\end{equation}
where $\eta(\bm{\theta})$ is a sampling operator that keeps information only at the position  $\bm{\theta}_i$, given by
\begin{equation}
\eta(\bm{\theta}) = \frac{A}{N_s}  \sum_{i=1}^{Ns} \delta(\bm{\theta}-\bm{\theta}_i).
\label{eq:eta}
\end{equation}
Here $\delta$ is the Dirac delta function and $N_s$ is the total number of stars over an area $A$. Equation \ref{eq:dis_samp} also contains measurement errors - $\varepsilon$, with $\langle \varepsilon\rangle =0$ and $\langle \varepsilon^2\rangle =\sigma^2_{\varepsilon\varepsilon}$, which is also sampled at the positions of the stars. With a representation of the observed field, we can now study the statistical properties of the field, namely the two-point correlation function of a field $f$,
\begin{equation}
\xi(\bm{\phi}) = \langle f(\bm{\theta})f(\bm{\theta}+\bm{\phi}) \rangle = \frac{1}{2\pi}\int d \bm{\ell} ~|\tilde{f}(\bm{\ell})|^{2} ~e^{-\imath\bm{\ell}.\bm{\phi}},
\label{eq:corr_func}
\end{equation}  
which is the well known result that the correlation function is the Fourier transform of the power spectrum $C^{ff}(\bm{\ell}) = |\tilde{f}(\bm{\ell})|^{2}$. From Equation \ref{eq:dis_samp} we can define the functions $g(\bm{\theta})= f(\bm{\theta})\eta(\bm{\theta})$ and $h(\bm{\theta})= \varepsilon(\bm{\theta})\eta(\bm{\theta})$. This allows us to decompose the observed power spectrum, $C^{obs}(\bm{\ell})$, into three components, $C^{gg}(\bm{\ell})$, $C^{hh}(\bm{\ell})$ and $C^{gh}(\bm{\ell})$ that are, respectively, the power spectrum from the autocorrelation of $g(\bm{\theta})$, the autocorrelation of $h(\bm{\theta})$ and the crosscorrelation of $g(\bm{\theta})$ and $h(\bm{\theta})$,
\begin{equation}
C^{obs}(\bm{\ell})=C^{gg}(\bm{\ell})+C^{hh}(\bm{\ell})+2C^{gh}(\bm{\ell}).
\label{eq:pk_obs}
\end{equation}
The power spectrum $C^{gg}$ is constructed from the Fourier space expression of $g(\bm{\theta})$,
\begin{equation}
\tilde{g}(\bm{\ell})=\frac{1}{(2\pi)^2}\int d\bm{\ell^\prime} \tilde{f}(\bm{\ell^\prime})\tilde{\eta}(\bm{\ell - \ell^\prime}),
\label{eq:fg}
\end{equation}
using the fact that 
\begin{equation}
\eta(\bm{\ell}) = \frac{A}{N_s} \sum_{i=1}^{N_s} e^{\imath \bm{\ell}.\bm{\theta}_i}.
\label{eq:feta}
\end{equation}
The power spectrum from autocorrelation of $g(\bm{\theta})$ is
\begin{equation}
C^{gg}(\bm{\ell}) = C^{ff}(\bm{\ell}) + \frac{1}{n_{s}(2\pi)^2} \int d\bm{\ell^\prime}C^{ff}(\bm{\ell^\prime}),
\label{eq:pgg}
\end{equation}
where $n_s=N_s/A$ is the star density. The integral in Equation \ref{eq:pgg} can be expressed in terms of the point variance of $g(\bm{\theta})$,
\begin{equation}
\sigma^{2}_{ff} = \langle |f(\bm{\theta})|^2\rangle = \frac{1}{(2\pi)^2} \int d\bm{\ell} C^{ff}(\bm{\ell}),
\label{eq:var}
\end{equation}
This can be substituted into Equation \ref{eq:pgg} to give
\begin{equation}
C^{gg}(\bm{\ell}) = C^{ff}(\bm{\ell}) + \frac{\sigma^{2}_{ff}}{n_s}.
\label{eq:bgg2}
\end{equation}
Applying the same procedure to the rest of the terms in Equation \ref{eq:pk_obs} leads to the expression,
\begin{equation}
C^{obs}(\bm{\ell})= C^{ff}(\bm{\ell}) +  C^{\varepsilon\varepsilon}(\bm{\ell}) +  2C^{f\varepsilon}(\bm{\ell}) + \frac{\sigma^{2}_{ff}}{n_s} +\frac{\sigma^{2}_{\varepsilon\varepsilon}}{n_s} + 2\frac{\sigma^{2}_{f\varepsilon}}{n_s}.
\label{eq:pk_obs2}
\end{equation}

We see that our observed correlation function is composed of three terms that come from the correlations and cross-correlations of the underlying field of interest and the errors, and three terms that look like white noise (scale independent) terms.

\section{Wiener Filtering}

The expression shown in Equation \ref{eq:pk_obs2} gives the raw power spectrum that would be measured. In practice, we can construct an estimator of the field $f$ that is better behaved.  Here we will do this using Wiener filtering, which is known to be an optimal filter for reducing the $\chi^2$ for the residuals. The estimator $\tilde{f}$ can be construct as
\begin{equation}
\widehat{\tilde{f}}(\bm{\ell}) = \Phi(\ell)\tilde{f}_{obs}(\bm{\ell}),
\label{eq:filter}
\end{equation}
where the filter $\Phi$ is given by
\begin{equation}
\Phi(\ell)=\frac{C^{ff}(\ell)}{C^{obs}(\ell)} = \frac{C^{ff}(\ell)}{C^{ff}(\ell) + C^{n}(\ell) },
\label{eq:wiener}
\end{equation}
with the observable being defined as $C^{obs}(\bm{\ell}) = C^{ff}(\bm{\ell}) +  C^{n}(\bm{\ell})$ and $C^{n}(\ell)$ is the power spectrum of the noise (see equation \ref{eq:pk_obs2}). From this we can calculate the residuals,
\begin{equation}
\widetilde{\delta f}(\bm{\ell}) = \widehat{\tilde{f}}(\bm{\ell}) - \tilde{f}(\bm{\ell}) = \Phi(\ell) \tilde{f}_{obs}(\bm{\ell}) -\tilde{f}(\bm{\ell}),
\label{eq:res}
\end{equation}
and the power spectrum of the residuals,
\begin{equation}
C^{\delta f \delta f}(\bm{\ell}) = \Phi^2(\ell) C^{obs}(\bm{\ell}) + C^{ff}(\bm{\ell}) - 2\Phi(\ell)C^{obsf}(\bm{\ell}).
\label{eq:pdfdf}
\end{equation}

If we assume that the measurement error, $\varepsilon(\bm{\ell})$, is not correlated to the signal, $f(\bm{\ell})$, in Equation \ref{eq:dis_samp} then the term for the observed power spectrum in Equation \ref{eq:pk_obs2} can be reduced to,
\begin{eqnarray}
C^{obs}(\bm{\ell}) & = & C^{ff}(\bm{\ell}) +  C^{\varepsilon\varepsilon}(\bm{\ell}) +  \frac{\sigma^{2}_{ff}}{n_s} +\frac{\sigma^{2}_{\varepsilon\varepsilon}}{n_s} \\
\label{eq:cscn}
 & = & C^{ff}(\bm{\ell}) +  C^{n}(\bm{\ell}) ,
\end{eqnarray}
where $C^{n}(\bm{\ell})$ contains the noise terms, which in this case are $C^{n}(\bm{\ell})=C^{\varepsilon\epsilon}(\bm{\ell}) +  \sigma^{2}_{ff}/n_s +\sigma^{2}_{\varepsilon\varepsilon}/n_s$. In this case the power spectrum of the residual in Equation \ref{eq:pdfdf} reduces to
\begin{equation}
C^{\delta f \delta f}(\bm{\ell}) = \big(\Phi(\ell)-1\big)^2C^{ff}(\bm{\ell}) + \Phi^2(\ell)C^{n}(\bm{\ell}).
\label{eq:pdfdf2}
\end{equation}
For simplicity, in what follows we will consider the isotropic case where there is no preferred direction. In this case, the powerspectra become dependent only on $\ell$ and so $C(\ell) \equiv C(\bm{\ell})$.  In this isotropic case, we can substitute the definition of the Wiener filter in Equation \ref{eq:wiener} into Equation \ref{eq:pdfdf2}, which leads to 
\begin{equation}
C^{\delta f \delta f}(\ell)= \frac{C^{n}(\ell)\big(C^{ff}(\ell)\big)^2 + \big(C^{n}(k)\big)^2C^{ff}(\ell) }{\Big(C^{ff}(\ell) + C^{n}(\ell)\Big)^2}.
\label{eq:isotropic}
\end{equation}

\section{Link with cosmic shear}

In \cite{2008A&A...484...67P} we showed how errors in the measurement of PSF size and ellipticity propagate into errors on the measured shear. This can be expressed as 
\begin{equation}
\bm{\delta{\gamma}} = \bm{F}\delta_{R^2} + G\bm{\delta_{\epsilon}},
\end{equation}
where $\delta_{R^2}$ is the variation of the PSF size, $\delta_{R^2} = (R_p^2 - \langle R_p^2 \rangle)/ \langle R_p^2 \rangle$ and  similarly $\bm{\delta_\epsilon}$ is the variation in PSF ellipticity. $\bm{F}$ and G are given by:
\begin{equation}
\bm{F} = \frac{1}{P^\gamma} \bigg(\frac{R_p}{R_g}\bigg)^2 (\bm{\epsilon_g} - \bm{\epsilon_p}),
\end{equation}
and
\begin{equation}
G = -\frac{1}{P^\gamma} \bigg(\frac{R_p}{R_g}\bigg)^2.
\end{equation}
Here, $P^\gamma$ is the shear susceptibility factor, $R$ and $\epsilon$ are the Radius and ellipticity and the subscripts p and g stand for PSF and galaxy, respectively. 

For the rest of this paper, we will focus  on the case where on average the errors in the two component of shear ($\gamma_1 \rm ~and~ \gamma_2$ where $\bm{\gamma} = \gamma_1 +i \gamma_2$) are the same, and thus the power spectrum can be written in terms of one of the components,
\begin{equation}
C_\ell^{\mathbf{\delta\gamma}} = 2 C_\ell^{\delta\gamma_1}.
\end{equation}
One of the shear components (for instance the first one) is given by,
\begin{equation}
\delta\gamma_1 = F_1 \delta_{R^2} + G\delta_{\epsilon_1},
\end{equation}
which leads to a systematic power spectrum of
\begin{equation}
C_\ell^{\delta \gamma_1} = F_1^2 C_\ell^{\delta R^2} + G^2 C_\ell ^{\delta\epsilon_1}.
\end{equation}

Both $R^2$ and $\epsilon_1$ are fields that are constructed from the measurements of the stars. From Equation \ref{eq:isotropic}, we know that the post Wiener filtering power spectrum of these two fields is given by
\begin{equation}
C_\ell^{\delta \chi} = \frac{C_\ell^{n\chi} (C_\ell^{\chi})^2 + (C_\ell^{n\chi})^2 C_\ell^{\chi}}{C_\ell^{\chi} + C_\ell^{n\chi}},
\end{equation}
where $\chi$ can be either $R^2$ or $\epsilon_1$.
The systematic limit from \cite{2008MNRAS.391..228A} can then be expressed as 

\begin{equation}
\label{eq:2comp}
\sigma_{sys}^2 = 2 \sigma_{\gamma_1}^2 = 2(F_1^2 \sigma^2_{sysR^2} + G^2 \sigma_{sys\epsilon_1}^2)
\end{equation}
\begin{equation}
\sigma^2_{sys\chi} = \int \ell^2 C_\ell^{\delta\chi} d\ln\ell .
\end{equation}

\subsection{Illustrative Examples}
\label{sec:illex}
\begin{figure} 
\centering
\resizebox{0.95\columnwidth}{0.95\columnwidth}{\includegraphics{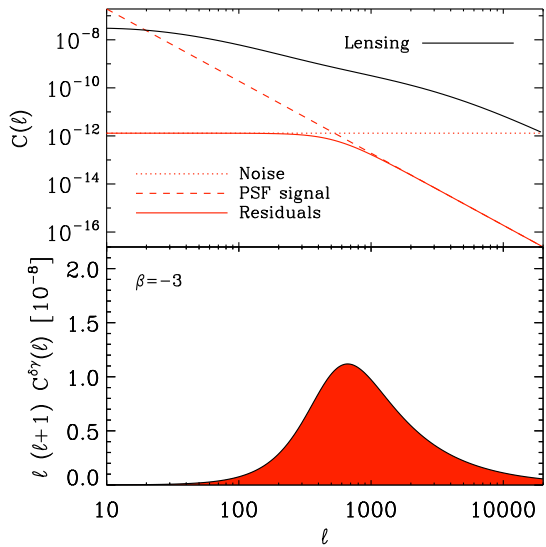}}
\caption{The upper panel shows a possible example of a PSF power spectrum. In this example, we focus on the impact of ellipticity variation (explicitly the variation in one of the components of ellipticity). For this example, we assume that both components of shear ($\gamma_1$ and $\gamma_2$) contribute equally to errors (as shown in Equation \ref{eq:2comp}). We have also chosen an example where the ellipticity variation and size variation contribute equally ($ F_1^2 \sigma^2_{sysR^2} = G^2 \sigma_{sys\epsilon_1}^2$).  With these assumptions, the example shown here leads to $\sigma_{sys}^2 = 10^{-7}$. The red dashed line shows the initial ellipticity power spectrum before Wiener filtering with $\beta=-3$. The dotted line shows the noise contributions ($C^{n}(\bm{\ell})$ of Equation \ref{eq:cscn}). The assumptions that go into calculating this noise level are given in Section \ref{sec:illex}. The solid red curve shows residual power spectrum after filtering.  We see that on large scales (small $\ell$), filtering (i.e. PSF fitting) is able to correct the original PSF to a large extent (down to the noise). On small scales (large $\ell$), the noise becomes dominant. Hence, the filtering process is not able to correct PSF, making us dependent on the underlying, inherent PSF variation.  The transition between these two regimes - in this example at $\ell \sim 500$ - marks the point where on larger scales the PSF correction depends on the quality of the analysis software and on smaller scales PSF variations need to be controlled in the observations themselves. The solid black curve is an example of the lensing power spectrum (see public code for details). We see that for $\beta = -3$ our requirement is that the pre-correction ellipticity power spectrum be roughly 1500 times smaller that the lensing power spectrum at $\ell_o =1000$ ($C_{\ell_o}^{lens} \sim 3 \times 10^{-10}$ and $C_{\ell_o}^{\epsilon_1}\sim 1.9 \times 10^{-13}$, which can also be expressed as $\ell_{o}^2C_{\ell_o}^{lens} \sim 3 \times 10^{-4}$ and $\ell_{o}^2C_{\ell_o}^{\epsilon_1}\sim 1.9 \times 10^{-7}$). 
The lower panel shows the post filtering residual power spectrum of the errors on the shear using a log-linear linear scale.  The integral of this curve (the red shaded area) corresponds to $\sigma_{sys}^2$ from this component.
}
\label{fig:fig1} 
\end{figure}

We now explore some example cases to illustrate how an analysis of the PSF variation can be set by a desired systematic target. For this we will make some simplifying assumptions.  First, we will assume that the systematic contributions from PSF size are comparable to those coming from ellipticity.  Explicitly, we set $\sigma_{sysR^2}^2 = \sigma_{sys\epsilon_1}^2$. We also set $\rm P_{\gamma} =1.84,~\langle\epsilon_g^2\rangle^{1/2} = 0.4,~\langle\epsilon_p^2\rangle^{1/2}=0.05 ~and ~R_g/R_p =1.5$ \citep{2009arXiv0901.3557P}. We also assume that the star density if 1 star per arcmin$^2$ with a S/N = 100, which in turn can be translated to errors on the PSF, per star, of $\sigma(\epsilon) = 3 \times 10^{-3}$ and $\sigma(R^2)/R^2 = 3 \times 10^{-3}$ (see \cite{2009arXiv0901.3557P} for justification). These are conservative assumptions for a Euclid like mission, but we should should note the our requirements will depend, moderately, on the choice of these parameters. Given this information, we can set requirements on the powerspectra of the PSF variation that are needed to reach a systematics level of $\sigma_{sys}^2 < 10^{-7}$, given a particular functional form. For simplicity we begin with a power-law
\begin{equation}
\label{eq:powlaw}
C_\ell^\chi = C^{\chi}_{\ell_o} \Bigg(\frac{\ell}{\ell_o} \Bigg)^\beta ,
\end{equation}
where $\ell_o$ is some convenient reference scale, we pick $\ell_o = 1000$. Note that the choice of $\ell_o$ has no impact on the results, for instance those shown in figure \ref{fig:fig1}. This choice only chances the $C^\chi_{\ell_o}$  that we quote. $C^\chi_{\ell_o}$ is the amplitude of the power spectrum at this $\ell$ value and $\beta$ is the slope of the power-law. 
In Figure \ref{fig:fig1} we see that the most relaxed requirements for $C^\chi_{\ell_o}$ come at roughly $\beta = -3$. This is because for slopes shallower than this, the contribution to that variance from small scales (high $\ell$) begins to dominate, whereas for steep spectra the variations on large scales start to become a problem.  At $\beta = -3$ we find that out requirement is that the pre-correction ellipticity power spectrum must be roughly 1500 times smaller that the lensing power spectrum at $\ell_o =1000$ ($C_{\ell_o}^{lens} \sim 3 \times 10^{-10}$ and $C_{\ell_o}^{\epsilon_1}\sim 1.9 \times 10^{-13}$. This can also be expressed as $\ell_{o}^2C_{\ell_o}^{lens} \sim 3 \times 10^{-4}$ and $\ell_{o}^2C_{\ell_o}^{\epsilon_1}\sim 1.9 \times 10^{-7}$). 

\begin{figure} 
\centering
\resizebox{0.95\columnwidth}{0.95\columnwidth}{\includegraphics{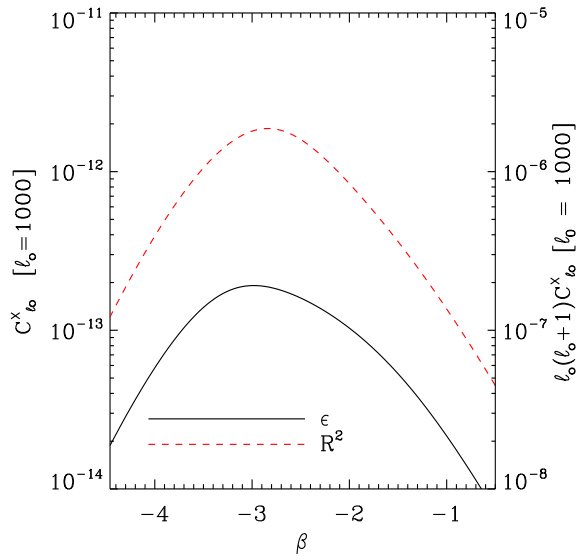}}
\caption{The tolerance on the powerspectra of ellipticity (solid black) and radius (red dashed) as a function of the power-law slope. We make the same assumptions about the relative strengths of each component as we do for Figure 1. To zeroth order the ratio of the powerspectra is given by the ratio $F^2_1/G^2$. We see that for both very steep and shallow values of beta, the requirements become more stringent. The most relaxed requirements are for $\beta \sim -3$ (shown in Figure \ref{fig:fig1}).}
\label{fig:fig2} 
\end{figure}

It is, therefore, interesting to consider an extension beyond the simple power-law. We investigate the limits that we set on a power spectrum with the following form,
\begin{equation}
C_\ell^\chi = C^\chi_{\ell_o} \Bigg(\frac{\ell}{\ell_o} +1 \Bigg)^{\beta_2 -\beta_1} \Bigg(\frac{\ell}{\ell_o} \Bigg)^{\beta_1} ,
\label{eq:cl2slope}
\end{equation}
where $\beta_1$ is the low $\ell$ slope and $\beta_2$ is the high $\ell$ slope. The results for this are shown in Figure \ref{fig:fig2}.  We also note that the diagonal that runs from [0,0] to [-4,-4], i.e. $\beta_1 = \beta_2 = \beta$ corresponds to the power law case shown in Figure \ref{fig:fig1}.

 \begin{figure} 
\centering
\resizebox{0.95\columnwidth}{0.95\columnwidth}{\includegraphics{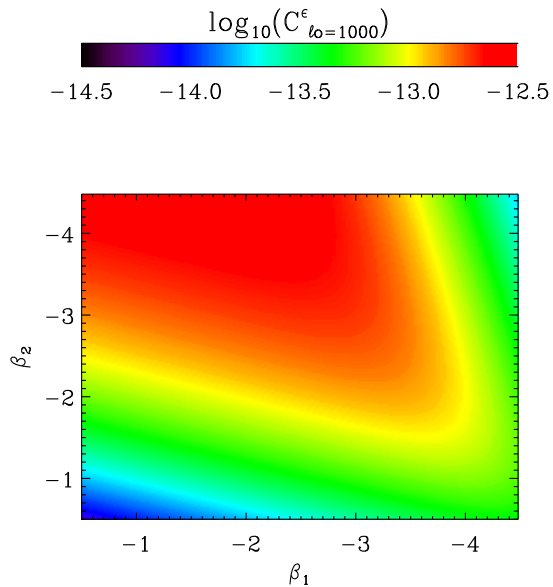}}
\caption{The tolerance on ellipticity power spectrum amplitude as a function of $\beta_1$ and $\beta_2$, the inner and outer slopes of Equation \ref{eq:cl2slope}. We see that the best results are given for spectra that are shallow for low $\ell$ and steep for high $\ell$. A diagonal cut through this figure (bottom left to top right corner) leads to results shown for the solid curve of Figure \ref{fig:fig2}. }
\label{fig:fig3} 
\end{figure}

Not surprisingly, we see that the most relaxed constraints on $C^\chi_{\ell_o}$ come from large $\beta_{1}$ (less negative) and small $\beta_{2}$ (more negative).
 
\section{Interpretation and Conclusions} 

We have investigated the way that the intrinsic variation of the PSF of an observation can be corrected using stars, which give a noisy estimate of the PSF at random points in the field. This allows us to asses the potential systematic contamination that comes from an observation with given ellipticity and size powerspectra for the PSF.  We show how software can be used on large scales to correct the PSF using the information form the stars. We have used Wiener filtering to model the PSF variation. For a Gaussian Random field this is known to be the optimal filter for reducing the residuals. On small scales however, our ability to calibrate the PSF fails due to the finite number of stars available. In this regime, lensing measurements become dependent on the underlying PSF variation of the instrument. The burden on these scales then falls on the hardware, since we need to perform measurements that have little small scale PSF power. This, therefore, provides us with a mechanism to make robust predictions about the systematic floor of lensing surveys, both current and future.

To illustrate the power of this approach, we place requirements on the size and ellipticity powerspectra that a Euclid-like survey would need to stay within the systematic requirement of $\sigma_{sys}^2 < 10^{-7}$ \citep{2008MNRAS.391..228A,2008arXiv0812.1966K}. We do this for the case where the errors on the two components of shear are comparable and the contributions to the final systematics from size and ellipticity are comparable. Ideally, the PSF of the observations would have a shallow slope on large scales ( $\beta_1 > -3$), which helps reduce the contributions to sample variance, and a steep small scale slope ($\beta_2 < -3$), which helps contain shot noise contributions. For the case where the initial PSF is a power-law with $\beta = -3$ (see Equation \ref{eq:powlaw}), we find that the power spectrum of PSF ellipticity at $\ell = 1000$ needs to be $ C^{\epsilon_1} < 2 \times 10^{-13}$ and the power spectrum of size ($R^2$)  needs to be $C^{R^2} <2\times 10^{-12}$. This can be compared to the lensing power spectrum at this scale of $C^{lens} \sim 3\times10^{-10}$.  Stated another way, this means that the pre-correction ellipticity power-spectrum needs to be 1500 times smaller than the lensing signal (at $\ell =1000$) for a Euclid-like survey to be statics limited rather than systematics limited. In the case of space-based surveys, this can be seen as a requirements on the auto-correlation power spectrum of the time variable instrument PSF. 

For ground-based surveys, if the auto-correlation power spectrum of images is found to exceed our requirement, mitigation strategies need to be found since we have only limited control over the PSF pattern, which is mostly dominated by atmospheric effects. For a Gaussian random field Winer filtering will give the smallest residuals when modeling. This makes our predictions optimistic. However for a non-Gaussian field other technics, e.g. Principal Component Analysis (PCA) proposed by \cite{2008JCAP...01..003J}, may be able to do better than Wiener filtering and reach comparable results to Wiener filtering with a Gaussian field. This alone may not be sufficient to meet our systematic requirements since this does not tackle the problem of a high initial PSF power spectrum. Other techniques would then be needed. As an example of possible options, the shape of a galaxy from one image can be correlated with the shape of another galaxy in a different image. To set the requirements on this we would also need to consider our sensitivity to the cross-correlation power spectrum of the PSF between the two images. 

\section*{Acknowledgements}
We thank Tom Kitching for detailed  comments and feedback on an early draft. We also thank Sarah Bridle, Lisa Voigt, Alan Heavens and the rest of the Euclid Weak Lensing Working Group members for useful discussions. AA is supported by the Zwicky Fellowship at ETH Zurich. SPH is supported by the P2I program, contract number 102759.

\bibliographystyle{astron}
\bibliography{/Users/aamara/Work/Mypapers/mybib}

\end{document}